\documentclass[fleqn,10pt]{wlscirep}
\usepackage[utf8]{inputenc}
\usepackage[T1]{fontenc}
\usepackage{caption}
\usepackage{subcaption}
\usepackage{booktabs}
\usepackage{float}
\usepackage{parskip}
\usepackage{hyperref}
\usepackage{multirow}
\usepackage{tablefootnote}
\usepackage{threeparttable}
\usepackage{xcolor}

\newcommand\numcomp{681 }
\newcommand\numnews{5436 }

\newcommand\numposnews{1595}
\newcommand\numnegnews{798}
\newcommand\numneutnews{3043 }
\newcommand\postnewsperiod{20}
\newcommand\trainstep{30}

\begin{document}

\title{New drugs and stock market: how to predict pharma market reaction to clinical trial announcements}

\author[1,2,*]{Semen BUDENNYY}
\author[1]{Alexey KAZAKOV}
\author[1]{Elizaveta KOVTUN}
\author[3]{Leonid ZHUKOV}

\affil[1]{Sber AI Lab, Moscow}
\affil[2]{Artificial Intelligence Research Institute (AIRI), Moscow}
\affil[3]{Higher School of Economics University, Moscow}

\affil[*]{Corresponding author: sanbudenny@sberbank.ru}

\begin{abstract}
Pharmaceutical companies operate in a strictly regulated and highly risky environment in which a single slip can lead to serious financial implications. Accordingly, the announcements of clinical trial results tend to determine the future course of events, hence being closely monitored by the public. In this work, we provide statistical evidence for the result promulgation influence on the public pharma market value. Whereas most works focus on retrospective impact analysis, the present research aims to predict the numerical values of announcement-induced changes in stock prices. For this purpose, we develop a pipeline that includes a BERT-based model for extracting sentiment polarity of announcements, a Temporal Fusion Transformer for forecasting the expected return, a graph convolution network for capturing event relationships, and gradient boosting for predicting the price change. The challenge of the problem lies in inherently different patterns of responses to positive and negative announcements, reflected in a stronger and more pronounced reaction to the negative news. Moreover, such phenomenon as the drop in stocks after the positive announcements affirms the counterintuitiveness of the price behavior. Importantly, we discover two crucial factors that should be considered while working within a predictive framework. The first factor is the drug portfolio size of the company, indicating the greater susceptibility to an announcement in the case of small drug diversification. The second one is the network effect of the events related to the same company or nosology. All findings and insights are gained on the basis of one of the biggest FDA (the Food and Drug Administration) announcement datasets, consisting of \numnews clinical trial announcements from \numcomp companies over the last five years.
\end{abstract}

\maketitle

\begin{keywordname}
pharmaceutical, stock market, clinical trials, time series, event study, NLP, BERT, TFT, GNN
\end{keywordname}

\section{Introduction}
The relation between the company market value and company events attracted the research community's attention from the very moment it became technically possible to collect and merge the corresponding data. One of the pioneer works is considered to belong to Dolley, James Clay in 1933 \cite{dolley1933characteristics}. In practice, the study of events finds application in a wide range of socio-economic areas. In particular, the event studies are widely used within the interface between the law and economics to evaluate the impact of change in the regulatory environment on the company's market value. In legal liability, it is exploited to assess damages\cite{mackinlay1997event}. In investment companies, event studies are widely used to form an investment strategy. In recent years, the pharmaceutical sector has become one of the most discussed because of COVID-19. 

A biopharmaceutical company's market value depends on two factors: a current product portfolio of the company with the intellectual rights associated with it and a portfolio of potential new drugs. Potential new drugs and clinical trials related to them directly influence the further market value behavior. The biopharmaceutical industry keeps high investment risks in research and development (R\&D) due to high operation and capital costs, relatively high value of drug's time-to-market (up to 15 years \cite{matthews2016omics},\cite{golec2007financial}), full dependence on regulatory agencies, high uncertainty \cite{djulbegovic2001acknowledgment} in clinical trials, and ethical aspects \cite{muthuswamy2013ethical}. Consequently, the market is highly sensitive to clinical trial results\cite{chen2021economic, vedd2019FDA}. Besides, the late-phase clinical studies are the most complicated and expensive (Phase III takes up to 40\% of total R\&D costs \cite{reuters2012cmr}).

Clinical trials are supported with hypothesis testing to evaluate a treatment effect numerically: the null hypothesis states that there is no treatment effect on the chosen clinical endpoint, while the alternative hypothesis posits that there is at least some treatment effect of the test drug\cite{guidance2017multiple}. This means that the clinical trial results reflect the treatment effect of a new drug (or repurposing of an existing drug towards a new indication) in terms of statistical significance. Understanding market value behavior for an upcoming clinical trial announcement is crucial to hedge potential financial risks (instant depreciation of market value, loss of stakeholders' confidence, or even default).

Most clinical trials in drug development are assessed at three classes of endpoints. The primary endpoints consist of the outcomes (based on the drug's expected effects) that establish the drug's effectiveness and safety features to support regulatory action. The secondary endpoint must demonstrate additional effects after success on the primary endpoint. All other endpoints are defined as exploratory. The common goal of all endpoints is the evaluation of the drug effects. This process can include assessment of clinical events (e.g., overall survival, progression-free survival, stroke, pulmonary exacerbation), monitoring of patient's symptoms (e.g., pain, depression), or measurement of functions (e.g., ability to work). 

The results of clinical trials of new medicines are supplied through official (company announcements, announcements of regulatory agencies) and non-official (mass media, pharma analytic agencies) sources. The most representative and widely exploited official sources of clinical announcements are the Food and Drug Administration (FDA), the European Medicines Agency (EMA), and other regulatory agencies providing the status of drug research and validation for certain areas (The Medicines and Healthcare products Regulatory Agency (MHRA), The Ministry of Health, Labour and Welfare, etc.). 

The previous research\cite{hwang2013stock} covered such sources as EMA and specialized medical agencies. Perez-Rodriguez\cite{perez2012product} analyzed outliers in market indicators of several pharmaceutical companies and assigned the causes to such outliers based on research and development activities covering clinical trials, drug design, and scientific publications. Tomovic \cite{tomovic2012longtermpharm} compared the impact of the FDA and dividend payment announcements on the market prices. In turn, the study\cite{niederreiter2022impact} identified characteristics analysis, which was indispensable for explaining the bio-pharmaceutical market reaction to announcements on product innovation. A couple of recent papers were focused on market influence during the pandemic. Rouatbi\cite{rouatbi2021immunizing} discussed the effect that vaccines have on market volatility and compared them for different regions of the world. 
With growing interest in advanced statistical approaches (e.g., machine learning, deep learning, time series analysis), event study is becoming less retrospective, and more attempts are aimed at building forecast models that evaluate the future behavior of market value for upcoming events \cite{baker2020machine, samitas2020machine}. The forecast models may support investors' decisions and evaluate associate risk according to the previous statistics. Moreover, machine learning-powered systems can be used not only for financial goals. The authors\cite{ma2007network, zeigler2021network} investigated target diseases of FDA-approved drugs using network theory. Ridder\cite{de2005predicting} anticipated outcomes of Phase III in the drug development process based on data from Phase II.
In the work \cite{elkin2021predictive}, in particular, the proposed model predicted whether a clinical trial would be successfully completed or not.
Manem\cite{manem2018network} developed a new approach for working with clinical trials based on network science, which potentially would better capture the complex nature of the disease. The machine learning models are also widely used in prognosis, prevalence, and mortality analysis of COVID-19 \cite{lalmuanawma2020covid_rew,ingram2021covid_pred}.

Complex relations between news, drugs, and financial networks allow researchers to investigate the company and market operating principles on a system-wide level rather than case-specific. Wan\cite{wan2021sentiment} employed natural language processing approaches to clarify the impact of news sentiment in the network of companies on the financial market. It turned out that positive news on some companies also positively influenced their neighbors in the co-occurrence network. There\cite{lacasa2021beyond} was a new method proposed to assess network similarity, which may be an indicator of causal links. An early warning system was considered a practical application for understanding interconnections between announcements and market response. The design of such systems based on machine learning algorithms was discussed\cite{baker2020machine, samitas2020machine}. 

Our study is organized as follows. In the \hyperref[sec:contribution]{Contribution} section, we highlight novel ideas and major accomplishments that are introduced in our research. In the \hyperref[sec:methods]{Methods} section, we describe key components of our pipeline for establishing relations between pharmaceutical companies' market prices and clinical trial announcements. The experiments' results are given in the \hyperref[sec:results]{Results} section. Principal findings and limitations are emphasized in \hyperref[sec:discussion]{Discussion} section. We summarize the results of the work done in the \hyperref[sec:conclusions]{Conclusions} section.

\section{Contribution} \label{sec:contribution}

In this research, our goal is to forecast the market value change caused by releases on clinical trial results, considering the pharmaceutical industry's peculiarities. The main contribution of the study is as follows.

\begin{itemize}
    \item Providing statistical evidence on the reasonableness of numerical evaluation of announcement impact. The conducted statistical tests indicate a relationship between result announcements and price change. 
    \item Analysis of one of the biggest trial result announcement datasets from the pharma industry. We provide reliable data insights on the basis of \numnews announcements of \numcomp public companies.
    \item Development of a unified framework for efficient preprocessing of clinical announcement data. The framework includes parts responsible for defining the sentiment polarity of the announcement, extracting valuable features, and predicting the expected return.
    \item Obtaining high-quality prediction for price change range. The combination of gradient boosting (GB) classifier with graph convolution network (GCN) encourages capturing data interdependencies and allows getting the strong performance of classification with the total weighted area under the receiver operating characteristic curve (ROC AUC) greater than 0.7.
\end{itemize}

\section{Methods} \label{sec:methods}

\begin{figure}[ht!]
    \centering
    \includegraphics[width=0.99\linewidth]{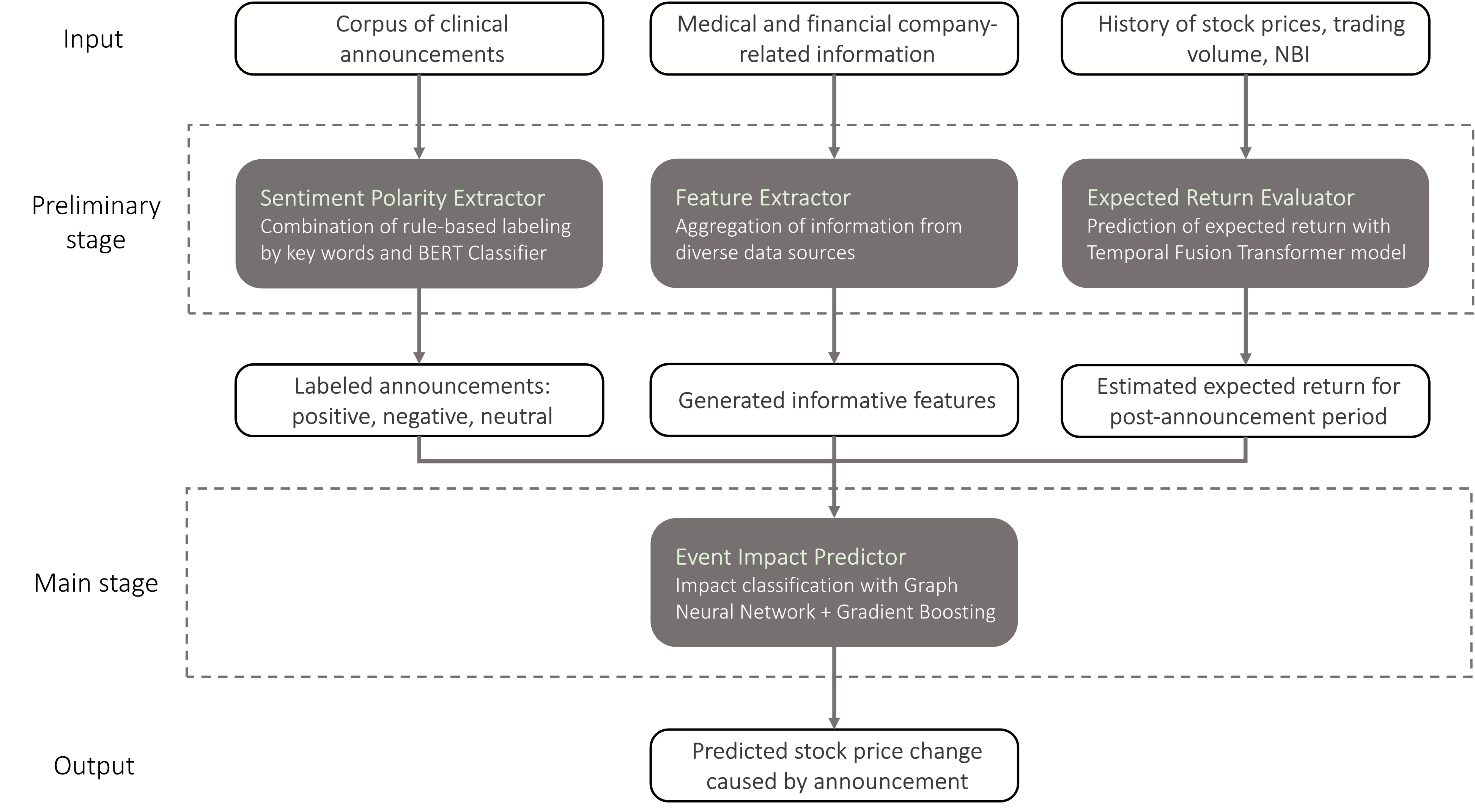}
    \caption{High-level pipeline for the prediction of stock price change induced by the clinical trial results. The preliminary stage consists of the data preparation, including the formation of the feature space and estimation of the expected return needed for the target variable calculation. The main stage includes the core classification model for the prediction of the price change range. } 
    \label{fig:tot_scheme}
\end{figure}

Due to relatively high costs of R\&D for drug discovery, pharma companies are highly sensitive to trial test announcements. In this section, we discuss the major components of the whole pipeline for change range prediction of company market value. The schematic representation of the pipeline is provided in Figure \ref{fig:tot_scheme}. There are two stages in the problem-solving process: preliminary and main. The preliminary stage serves as a data preprocessing. It focuses on the formation of representative feature space and the evaluation of the expected return, which is required to calculate the target variable. In particular, at the preliminary stage, we do the following. Firstly, we provide details on how we define the critical characteristic of the announcement, namely, its sentiment polarity. Secondly, we describe other generated attributes related to market, company, and announcement in general. The identified announcement polarity and other extracted features constitute the entire feature space, which we design for the training machine learning model. Thirdly, we explain the approach to calculating the expected return necessary for obtaining the target measure of announcement impact. The main stage consists of the core classification model that makes its prediction of the price change range based on the input feature space prepared at the preliminary stage. 

\begin{figure}[ht!]
    \centering
    \includegraphics[width=0.95\linewidth]{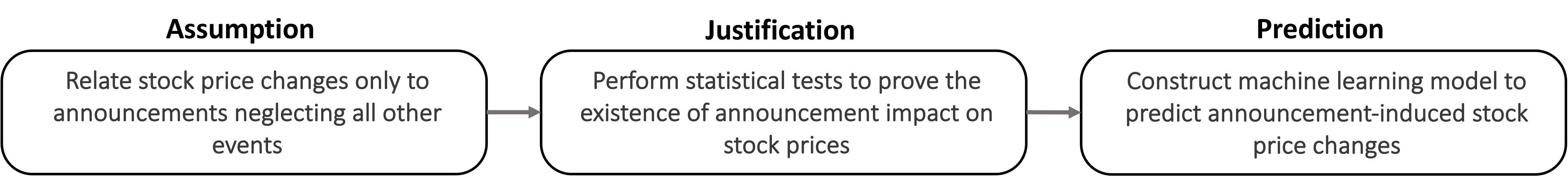}
    \caption{Visualization of logic for correlating the announcement events with the stock price changes. The essential points consist in making an assumption, performing statistical tests, and creating a predictive model.} 
    \label{fig:logic_scheme}
\end{figure}

It is important to note that we consider events related exclusively to the trial test promulgation, neglecting other events nearby in time. By far, near events could be essential (quarterly financial reports, news about merging and acquisitions, news on companies policy changes, new CEO announcement, etc.), but we have no information on them. Nevertheless, their effect is inherently present and can be noticed in the market price behavior. Further, we perform statistical tests to prove the impact of the trial test announcements on the company market value and follow the discussed pipeline for the influence estimation. The overall logic is demonstrated in Figure \ref{fig:logic_scheme}. More information on the limitations and assumptions adopted in our research is provided in the \hyperref[sec:discussion]{Discussion} section. The prediction of the clinical result itself is a self-contained complex task demanding a detailed investigation and is not examined in the frame of this work. 

\subsection{Sentiment polarity extraction from announcement}
As we aim to predict the market value change induced by an announcement, the identification of its sentiment polarity plays a pivotal role. We emphasize three polarity groups: the corpus of positive announcements (e.g., clinical trials approvals), the corpus of negative announcements (e.g., clinical trials termination, negative results), and finally the corpus of neutral announcements. To define the sentiment polarity, we process the collected corpus of historical releases in three phases:

\begin{enumerate}
    \item We compose the initial dictionaries with keywords, which reflect the announcement polarity. Examples of positive words are "approve", "meets", and "show". As indicators of the negative polarity, we set "halted", "no differentiation from placebo", "did not reach", and "failed". Then, we construct a rule-based announcement mark-up by checking for the presence of the mentioned words. 
    \item We take the pre-trained bidirectional encoder representations from transformers (BERT)\cite{devlin2018bert, devlin2018open} model and train it on all FDA trial result texts labeled with a rule-based method. After that, the trained BERT model is used to classify the announcements and reveal the examples on which it makes mistakes. Dictionaries are complemented with additional keywords extracted from mistakenly classified announcements.
    \item We create final announcement mark-up by leveraging updated dictionaries. The examples of words that are added with the help of the BERT model are "demonstrate", "potential", "accepted", "encouraging" for the positive case and "terminated", "discontinued", "insufficient", "paused" for the negative case.
\end{enumerate}

Thus, we use the power of the pre-trained BERT model as a support tool for more accurate mark-up. We deem that labeling announcements by keywords is a reasonable approach because the message of most announcements is delivered in a pronounced way. The obtained sentiment polarity serves as one of the features of the core classification model. 

\subsection{Aggregation of relevant information}

Following the common machine learning paradigm of data preprocessing, we want to construct a feature space that encompasses valuable information from various data sources to get a top-quality predictive solution. In our problem statement, the feature space is defined as a set of parameters that impact the stock price change after the announcement takes place. Hence, we generate new features that belong to one of the three domains: market, company, or announcement. Market and company features cover the trade and financial aspects of the company's operation. The announcement features incorporate sentiment polarity and medical-related attributes. The overall view of the constructed feature space is as follows.

\begin{itemize}
    \item Market features. These features describe the stock prices and index dynamics. Namely, they include the mean number of trading volume peaks per year, duration of the last trading volume peak, NASDAQ biotechnology index (NBI), and stock price trend for the last 30 days before the event and the previous 30-day trend (from 60 to 30 days before the event).
    \item Company features. We extract the company features from the annual reports. This does not consider quarter reports because not all companies publish them. We consider three types of reports to extract valuable information. The first type is the income statement, normalized by "Total Revenue". The second type of report is the balance sheet, values of which are normalized by "Cash From Operating Activities". Also, we take such features as "Total Common Shares Outstanding", "Full-Time Employees", and "Number of Common Shareholders" without further normalization. The third type is the cash flow. We normalize it by "Total Equity".    
    \item Announcement features. Each pharmaceutical announcement can be associated with a specific nosology. We get the International Classification of Diseases 10th Revision codes (ICD-10 codes) by matching the mentioned diseases in the announcement texts with the particular codes and use them as categorical features. Besides, the extracted sentiment polarity is also referred to as the announcement features. 
\end{itemize}

% \subsection*{Event impact evaluation}
\subsection{Target measure of event impact and evaluation of expected return}
The numerical effect of a certain event is usually examined as a difference between the actual return dynamics $R(t)$ and the expected returns $ER(t)$ within a post-event period, limited to $T$ days after the announcement is made ($t=0$). Thereby, the abnormal return $AR(t)$ on day $t$ after the announcement takes place can be defined as follows $AR(t)=R(t) - ER(t)$. To measure the integral effect within a given post-announcement period $(0,T]$, we use normalized cumulative abnormal returns (NCAR). NCAR is defined by the following formula: $ NCAR_T = \int_0^T AR(t) dt \Big/ \int_0^T ER(t) dt$, in which normalization is required to unify the target feature over announcements for various companies. $NCAR_{\postnewsperiod}$ is chosen as a target value we aim to predict for upcoming clinical releases. The motivation for taking the window width of the post-announcement period equal to $T=\postnewsperiod$ stems from the values of reacting period. More information on selecting the post-announcement period is given in Supplement. 

The expected return dynamics prediction is of high importance, as it directly contributes to the target value. The expected return is evaluated with temporal fusion transformer (TFT) model \cite{lim2021temporal}, comprised of 1 LSTM layer \cite{salman2018single} and 3 attention heads \cite{vaswani2017attention}. The model's architecture makes it possible to consider various aspects of trading without manually entering cause-and-effect relationships. We use mean absolute percent error (MAPE) as a loss function for our model because of its independence from the absolute stock price. The model is pre-trained on the whole trading history of the company excluding near announcement dates through the use of a window comprising {\trainstep} consecutive days for training and {\postnewsperiod} days for prediction. The positions of this window in the timeline are selected arbitrarily. After that, for each event, we train the model on 90 days of the trading history before the event date, shifting the same window by one day each time in order to allow the model to understand short market trends. The whole time series is normalized by the first value in it. 

The trading history we process with the TFT model consists of stock price, trading volume, and NBI. All of them allow estimating the expected return more precisely, taking into account implicit underlying factors. The use of trading volume facilitates accounting for information on hidden events, such as an announcement of dividends, quart reports, and many others. NBI is used as a global indicator in the pharma market. Within the NBI calculation, more than 100 companies are accounted for. This factor allows not to concentrate on one company but to evaluate the whole pharma market.

\subsection{The core classification model for price range prediction}

To predict the influence of FDA announcements on the stock price, we train the model that matches the input feature space composed of information on market, companies, and announcements with the target impact measure, $NCAR_{\postnewsperiod}$. In this work, the concepts of NCAR and price change are used interchangeably. Due to the insufficient number of regarded clinical announcements, a pure regression model for the stock prices does not provide acceptable quality. Therefore, we reduce the problem by transforming a regression setting into a classification. Instead of price change prediction, we predict the range of price change. The proposed problem statement that involves operating with the ranges grants knowledge of the price change's sign and amplitude. That is why the reformulated problem practically remains a significant concern in terms of risk evaluation. 

As a core classifier, we use an ensemble of algorithms consisting of GCN\cite{Kipf2017GCN} and GB\cite{Chen2016XGBoost}. At the start, we solve our classification problem using GCN exclusively. This model takes as input a graph of all announcement interconnections. There is an edge between two events in the graph if i) the earlier event happens with the same company or nosology as another one, and ii) the time period between events is less than one year. The architecture of GCN consists of 3 fully connected layers and 2 graph convolution layers\cite{ivanov2021boost}. Each node in the graph is represented as a vector that encapsulates the features of the corresponding announcement. The key idea behind the leveraged graph model is that the currently considered event updates its vector representation by exchanging the information between the neighboring nodes. 
% The example of a subgraph is presented in Figure \ref{fig:GILD_graph}. In this subgraph, the current announcement view is only affected by another event whether they are linked through the maximum of two edges. 
The motivation for taking the time period for establishing the event connections to be less than one year is in the optimal number of neighbors of the considered event, which is not too small to produce an uninformative neighborhood and not too big to cause event overlapping and muting. 
The purpose of GCN is to classify the price change caused by the announcement, so its output is the probabilities of belonging to a particular class. The classification is done proceeding from the resulting node representations. After dealing with the event graph, we leverage GB to get final class probabilities. Eventually, GB takes the composed feature space as input together with the output probabilities of the GCN model to construct the final prediction. Adopting the graph promotes the capturing of event interconnections and enhances the overall predictive quality.

\section{Results} \label{sec:results}

In this section, we review the results from each component of our overall pipeline for the event impact prediction. We start with the discussion of sources from which the dataset is gathered. Then, we describe the results of the sentiment polarity identification. After that, we provide outcomes of the statistical tests and data analysis related to the dependence of the stock price change behavior on the announcement polarity and company background. Also, we discuss the findings from the expected return evaluation. Finally, we introduce metrics related to our target problem of predicting stock price change ranges and feature importance evaluated by the Shapley additive explanations (SHAP) method. 

\subsection{Sentiment polarity evaluation}

Sentiment polarity is the characteristic of an announcement that predominantly determines the subsequent market response. We leverage the power of the pre-trained BERT model in our labeling process to make the final mark-up more reliable. First, we juxtapose the answers on sentiment polarity obtained from the rule-based approach with the initial keywords and the trained BERT model. Second, we analyze the differently labeled announcements and retrieve additional keywords from them to update the rule-based mark-up. The comparison of the number of divergences and coincidences in answers before and after the keyword update is provided in Table \ref{tab:labeling}. As we can see, the number of not matching answers of the rule-based method and the BERT model decreases after the keyword supplement. In addition, we observe a drop in the number of announcements labeled as neutral, since more announcements become emotionally charged as positive and negative. Eventually, the rule-based labeling and BERT predictions become more coherent, which leads to a higher quality of the mark-up. 

Originally, we deal with the following class distribution: \numposnews, \numnegnews, and \numneutnews of positive, negative, and neutral announcements, respectively. However, after the discussed procedure of sentiment polarity identification, we exclude all events related to private companies because of the inaccessibility of their stock prices. Thus, for further analysis, we have 1105 positive and 549 negative announcements at our disposal.

\begin{table}[h!]
\centering
\caption{Comparison of the sentiment polarity mark-ups derived from two methods, rule-based and trained BERT, before and after the keyword update. The number of divergences represents the number of announcements labeled differently by these methods. Whereas, for example, the number of coinciding positives reflects the number of announcements that are positive from the perspective of both methods.}
\begin{tabular}[t]{lcccccccccc}
\toprule
 & \# divergences & \# coinciding positives & \# coinciding negatives & \# coinciding neutrals \\
\midrule
With the initial keywords & 207 & 1447 & 445 & 3337 \\
With the updated keywords  & 66 & 1562 & 765 & 3043 \\
\bottomrule
\end{tabular}
\label{tab:labeling}
\end{table}

\subsection{Statistical tests for impact analysis of announcements}

\begin{figure}[h!]
\centering
\begin{subfigure}[t]{.5\columnwidth}
  \centering
  \includegraphics[width=\linewidth]{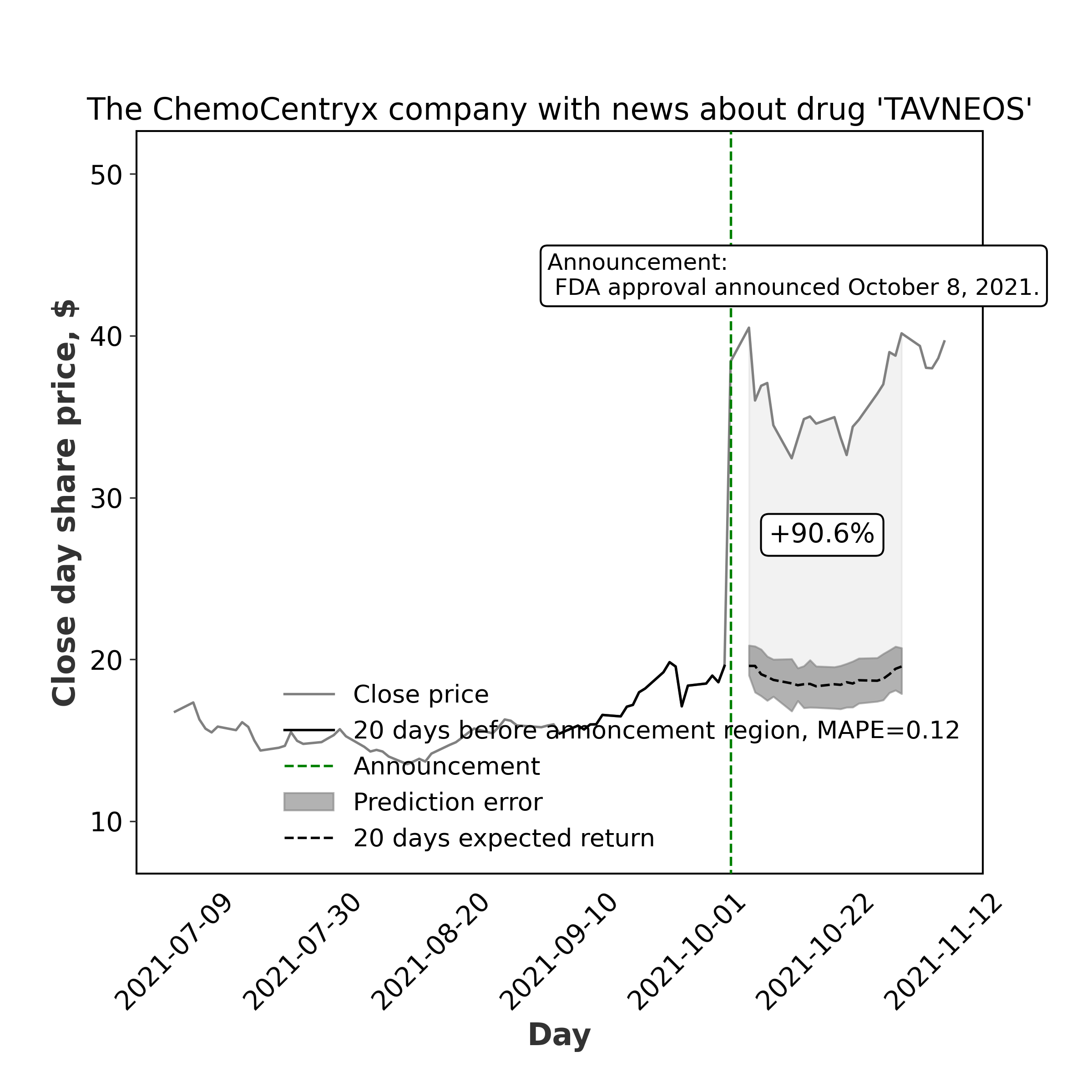}
  \caption{ChemoCentryx Announces FDA Approval of TAVNEOS™ (avacopan) in ANCA-Associated Vasculitis \cite{ChemoCentryx2021}}
  \label{fig:CCXI_TAVNEOS}
\end{subfigure}%
\begin{subfigure}[t]{.5\columnwidth}
  \centering
  \includegraphics[width=\linewidth]{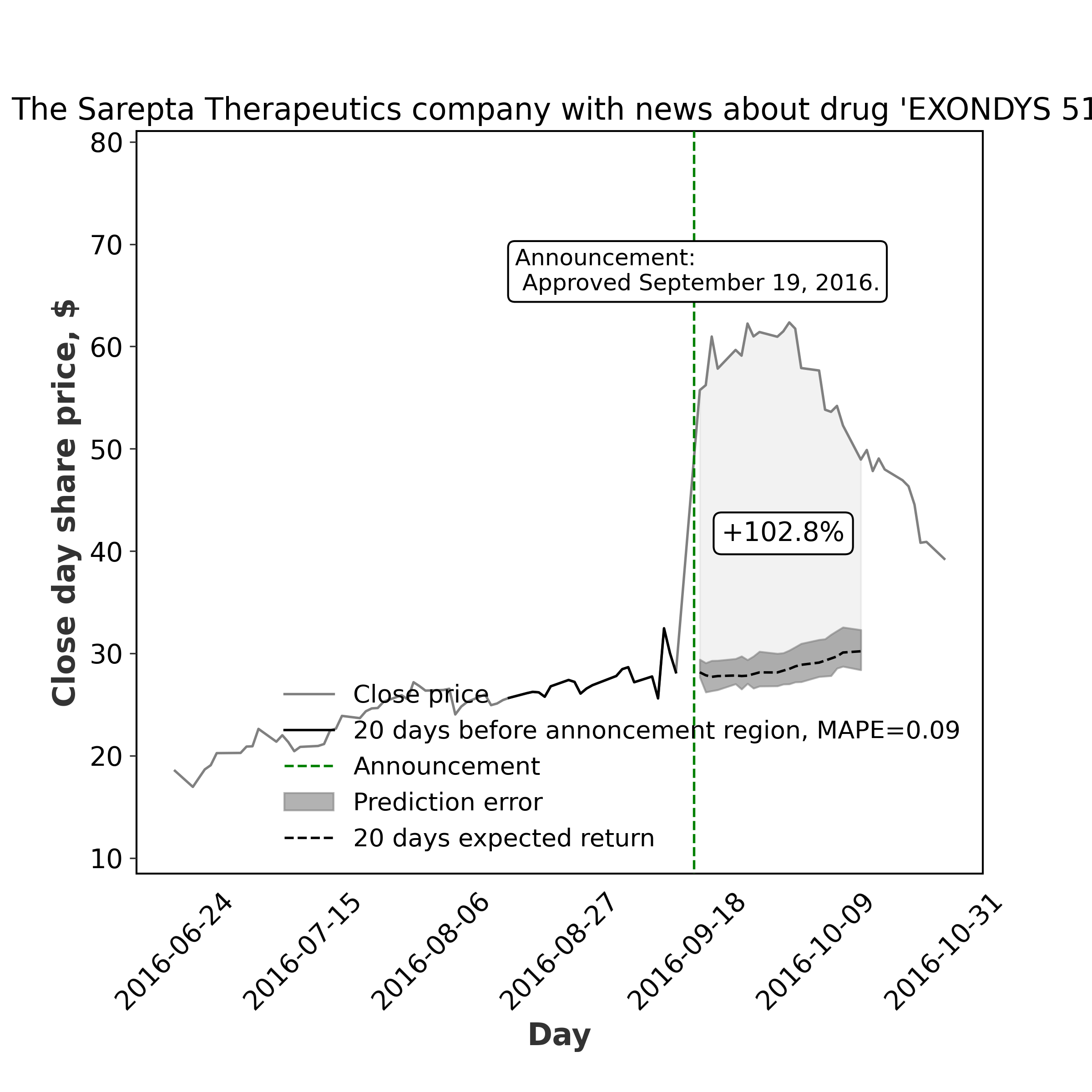}
  \caption{Sarepta Therapeutics announces FDA accelerated approval of EXONDYS 51 injection, an exon skipping therapy to treat duchenne muscular dystrophy patients amenable to skipping exon 51\cite{Sarepta2016}}
  \label{fig:SRPT_EXONDYS}
\end{subfigure}
\begin{subfigure}[t]{.5\columnwidth}
  \centering
  \includegraphics[width=\linewidth]{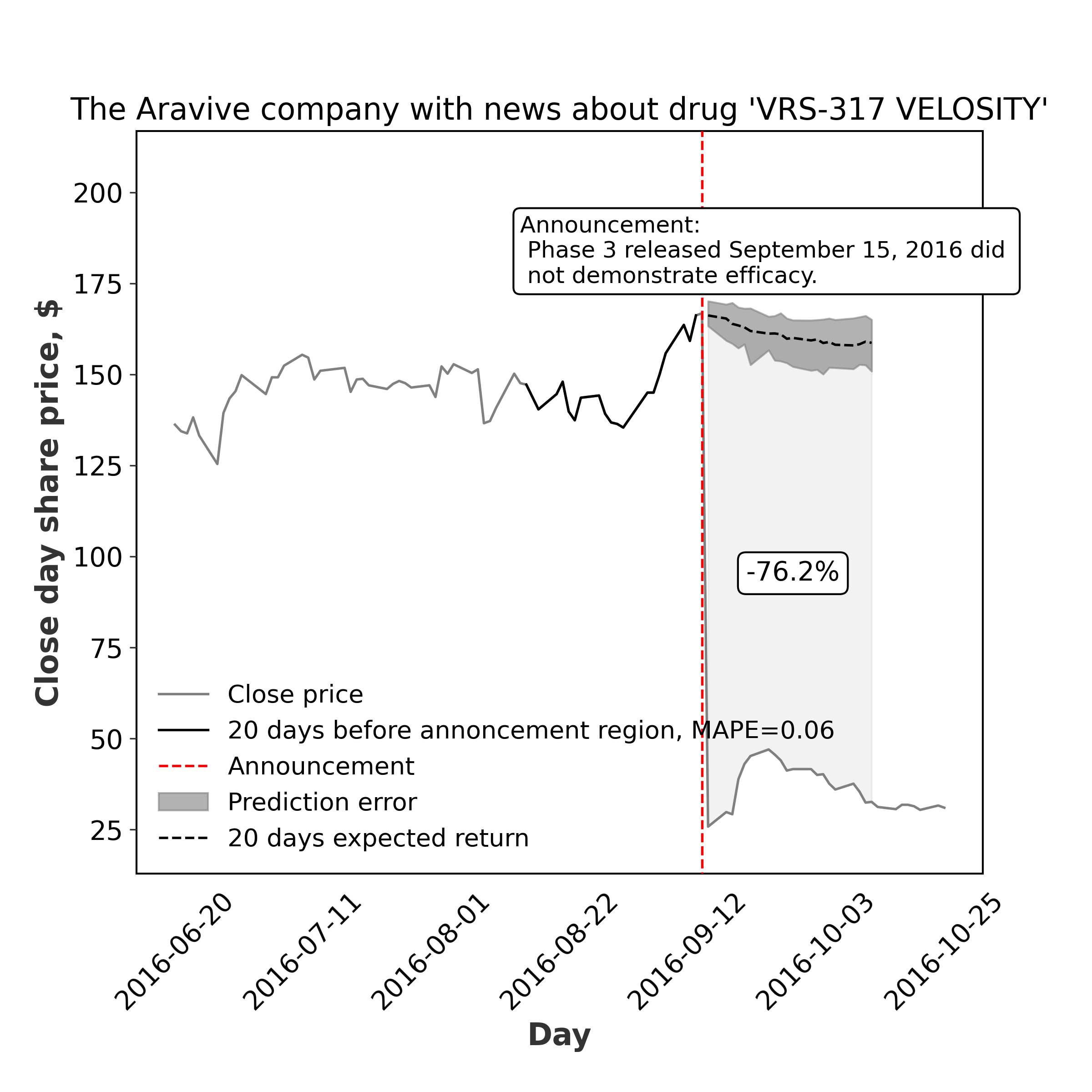}
  \caption{The Resolve™ trial, a Phase 3 trial of our RSV F Vaccine in 11,856 older adults (60 years of age and older), did not meet the pre-specified primary or the secondary efficacy objectives, and \\ did not demonstrate vaccine efficacy\cite{Novavax2016}}
  \label{fig:FBRX_FB_401}
\end{subfigure}%
\begin{subfigure}[t]{.5\columnwidth}
  \centering
  \includegraphics[width=\linewidth]{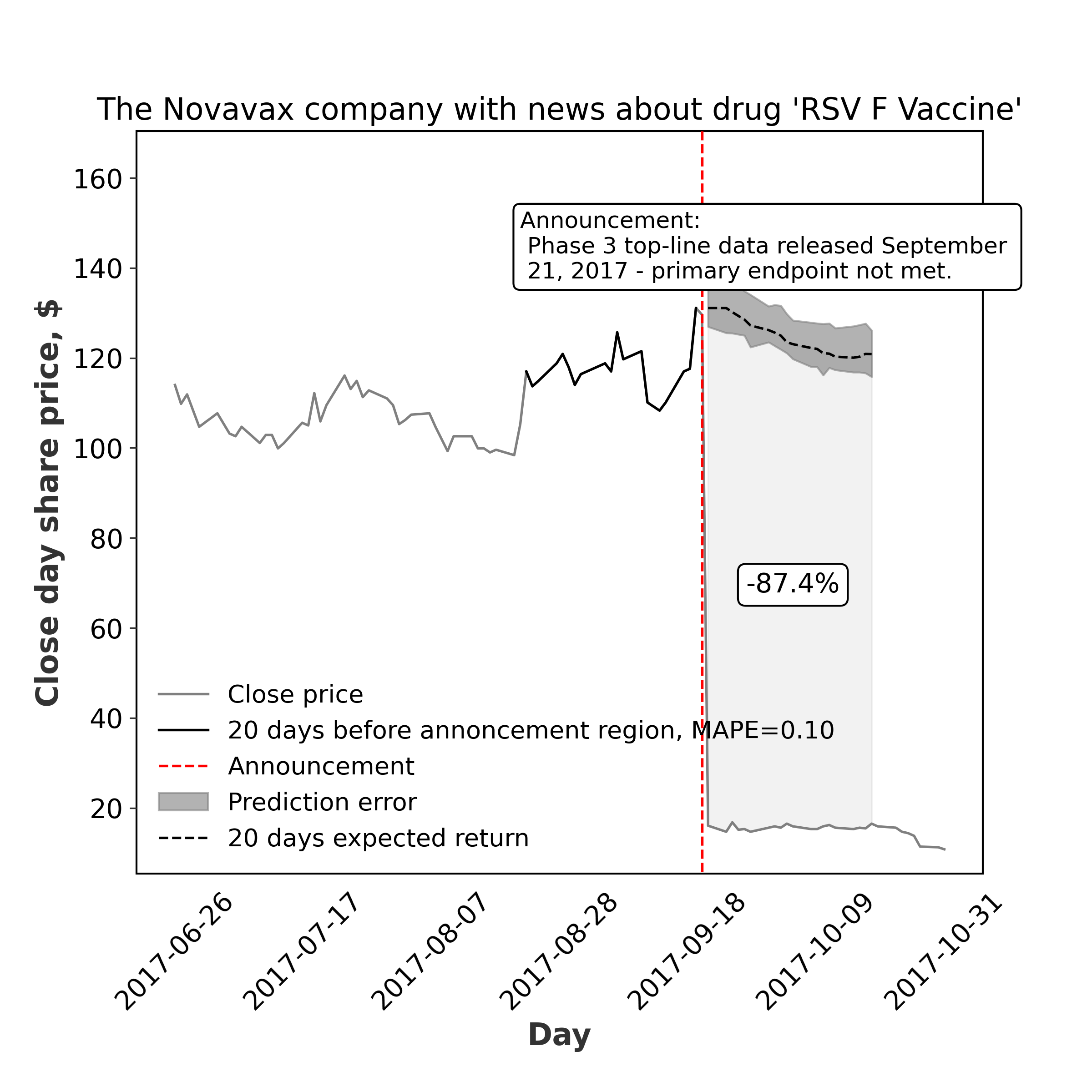}
  \caption{Announced that the VELOCITY Phase 3 clinical trial of somavaratan in pediatric growth hormone deficiency (GHD) did not meet its primary endpoint of non-inferiority\cite{Aravive2017}}
  \label{fig:ARAV_VRS_317_VELOC}
\end{subfigure}
\caption{Representative examples of the stock price changes according to the FDA trial clinical results (both positive (\ref{fig:CCXI_TAVNEOS}, \ref{fig:SRPT_EXONDYS}) and negative (\ref{fig:FBRX_FB_401}, \ref{fig:ARAV_VRS_317_VELOC}) announcements are examined).}
\label{fig:limit_cases}
\end{figure}

In some cases, the clinical results announcements can have an extreme effect on the company's financial position. The examples of the most influential clinical events, accompanied by the significant stock price change, are demonstrated in Figure \ref{fig:limit_cases}. 

\begin{figure}[h!]
    % \centering
    % \raggedright
    % width=1.1\linewidth
    \includegraphics[width=1\linewidth]{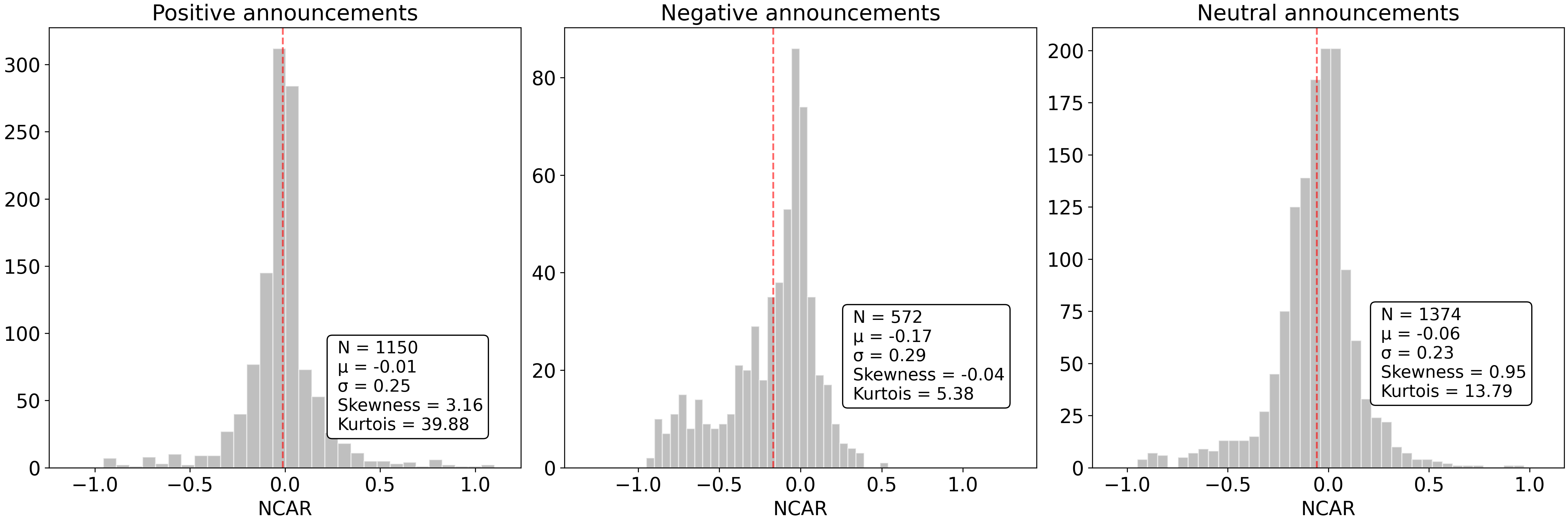} 
    \caption{The price change distributions ($NCAR_{\postnewsperiod}$) after the FDA announcement with positive, negative, and neutral contents takes place. The statistical parameters are provided in the boxes (number of events $N$, mean value $\mu$, standard deviation $\sigma$, skewness, and kurtois coefficients). The mean values are depicted with red dashed lines.}
    \label{fig:distributions_3sub}
\end{figure}

Our metric for the cumulative characteristic of the price change is $NCAR_{\postnewsperiod}$. The obtained $NCAR_{\postnewsperiod}$ distributions for the announcements of different sentiment polarities are shown in Figure \ref{fig:distributions_3sub}. For each announcement group, the Kolmogorov–Smirnov normality test with a p-value of 0.05 is performed. As a result, we get a p-value equal to $7 \cdot 10^{-9}$ for the negative announcements, a p-value equal to $ 10^{-34}$ for the positive announcements, and a p-value equal to $3 \cdot 10^{-16}$ for the neutral announcements, which indicates an abnormality of distributions. Indeed, it can be seen from Figure \ref{fig:distributions_3sub} that there are the abnormality and asymmetry in distributions related to the considered polarity groups. Meanwhile, we observe a significant number of negative announcements with positive $NCAR_{\postnewsperiod}$ values, and vice versa. This mismatch indicates that, for instance, positive news (e.g., Phase III acceptance) does not necessarily imply a positive impact on the stock prices. The negative announcements have a rather less mismatch ratio (most of the negative announcements are indeed connected with the negative stock price changes). The inconsistency between the announcement sentiment polarity and actual price change can be explained by the presence of other contributing factors (e.g., financial reports, merging or acquisition announcements, etc.) that are not considered in the frame of this research. A deeper inconsistency analysis is provided in Supplement.

Upon the disclosure of the non-normality of events distributions, we perform a rank non-parametrical statistical Mann–Whitney U test with a p-value of 0.05 to compare the positive and negative announcement distributions with non-announcement distributions. The non-announcement samples are generated from {\postnewsperiod}-day period before the announcement. The null hypothesis in this test states that positive and negative distributions do not differ from the non-announcement one. The p-value in the U test for comparing positive and non-announcement distributions is equal to 0.34. This means that the null hypothesis can not be rejected. In the U test with negative and non-announcement distributions, the obtained p-value is equal to $2 \cdot 10^{-13}$. In this case, the null hypothesis can be rejected.    

Therefore, it is statistically proven that there is an influence of the announcements on the stock prices. This fact confirms the validity of our problem statement, in which price changes correlate with the announcements of the trial results. Meantime, the analyzed distributions disclose that the responses to the positive and negative announcements are inherently different, as well as the sign of the price change effect can not be straightforwardly determined from the announcement polarity. 

\subsection{Impact analysis of company background on stock prices}

\begin{figure}[h!]
    \centering
    \includegraphics[width=\linewidth]{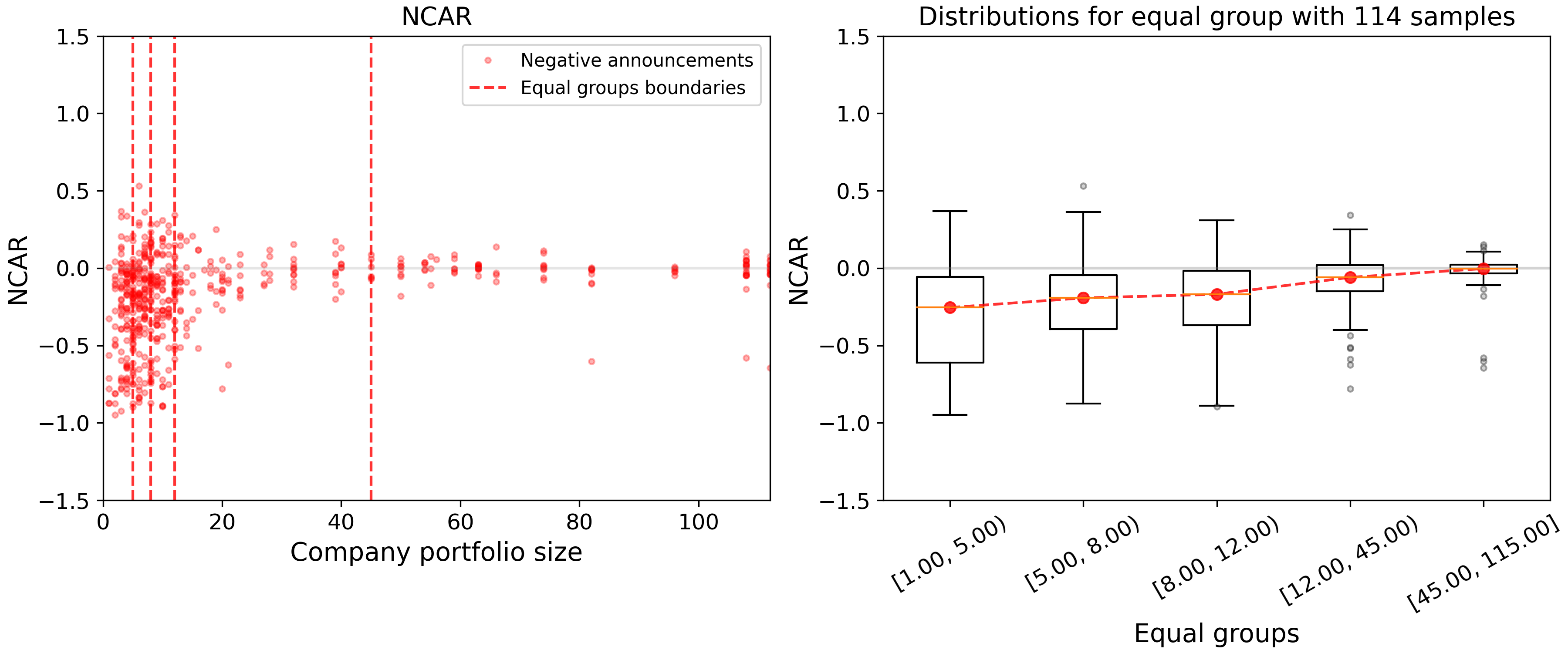}
    \caption{Dependence of the price changes over equal groups on size of the company drug portfolio for the negative announcements. In the left part, the red dashed lines divide announcements into equal groups (with the same number of announcements inside). In the right part, the red dashed line goes through the median values of each group's stock price changes.} 
    \label{fig:portfolio_size_neg}
\end{figure}

\begin{figure}[h!]
    \centering
    \includegraphics[width=\linewidth]{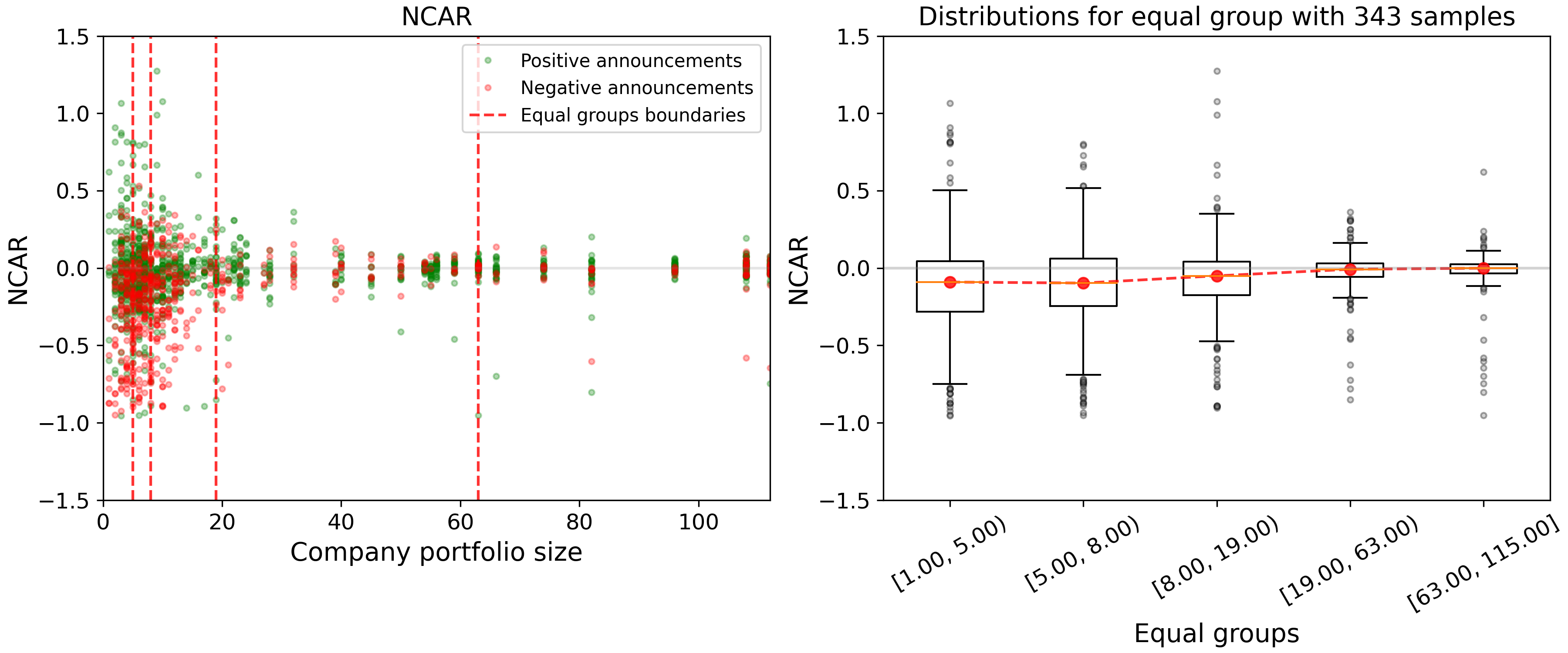}
    \caption{Dependence of price change over equal groups on the company's drugs portfolio size for negative and positive announcements. In the left part, the red dashed lines divide announcements into equal groups (with the same number of announcements inside). In the right part, the red dashed line goes through the median values of each group's stock price changes.}
    \label{fig:portfolio_size}
\end{figure}

One of the key characteristics of the pharmaceutical company's background is its drug portfolio size. This feature indicates the number of products that are supplied to the market. The portfolio size represents company scope and capacity, depending on which the reaction to the announcements of the same polarity may differ. In Figure \ref{fig:portfolio_size_neg} and \ref{fig:portfolio_size}, the price change is given for the various number of products in the portfolios of the examined companies. This demonstrates that small companies are more susceptible to any announcements in terms of their stock prices, whereas the positions of big companies are more robust. Moreover, for small companies, it is more probable to get a negative stock price change as a reaction to some event. More information on the dependence of stock price change on the company characteristics is given in Supplement. To sum up, we observe that the patterns of price changes can depend significantly on the company's background, which is the strong motivation for the generation of diverse features for the predictive model. 

\subsection{Expected return evaluation results}

One of the essential steps is an evaluation of the expected return, which is necessary for the calculation of an event's impact on stock prices. The time series is handled using the TFT model. To get a sense of its performance within our problem setting, we estimate the mean value of errors on {\postnewsperiod}-day periods before events. To do so, we train TFT on data from 110 days to {\postnewsperiod} days before events using the same sliding window described in the \hyperref[sec:methods]{Methods} section. After the model training phase, we predict stock prices for {\postnewsperiod}-day periods before events. We obtain MAPE of 0.07 for the dataset with positive and negative events. To conclude, the TFT model shows the good predictive quality and can be used for the appropriate evaluation of the expected return in a post-event period.

\subsection{Results on classification task for price change} 

\begin{figure}[h!]
    \begin{subfigure}[t]{0.5\columnwidth}
        \centering
        \includegraphics[width=0.98\linewidth]{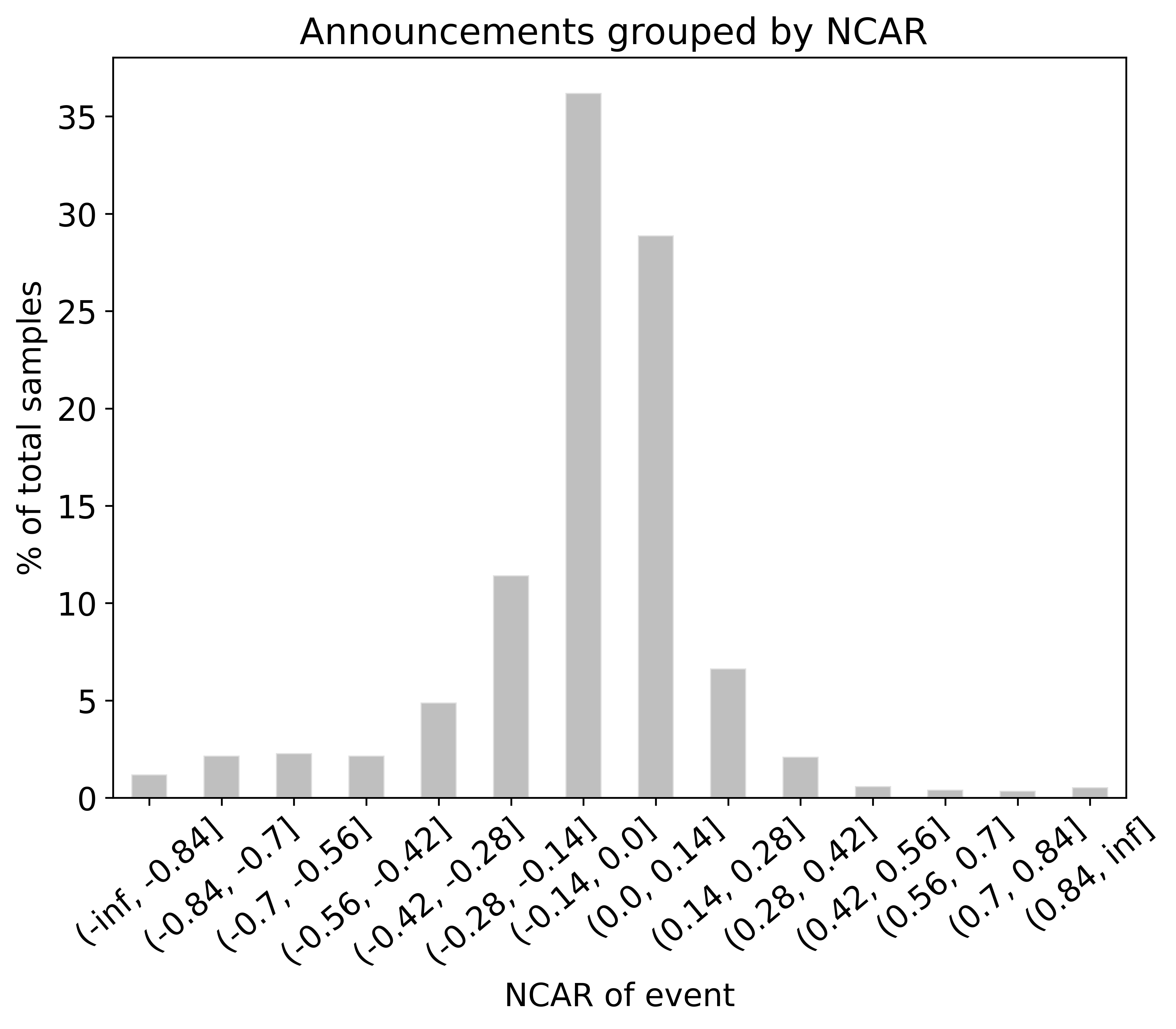}
        \caption{Distribution of events grouped by values of stock price change.} \label{fig:amp_distribution}
    \end{subfigure}
    \begin{subfigure}[t]{0.5\columnwidth}
        \centering
        \includegraphics[width=\linewidth]{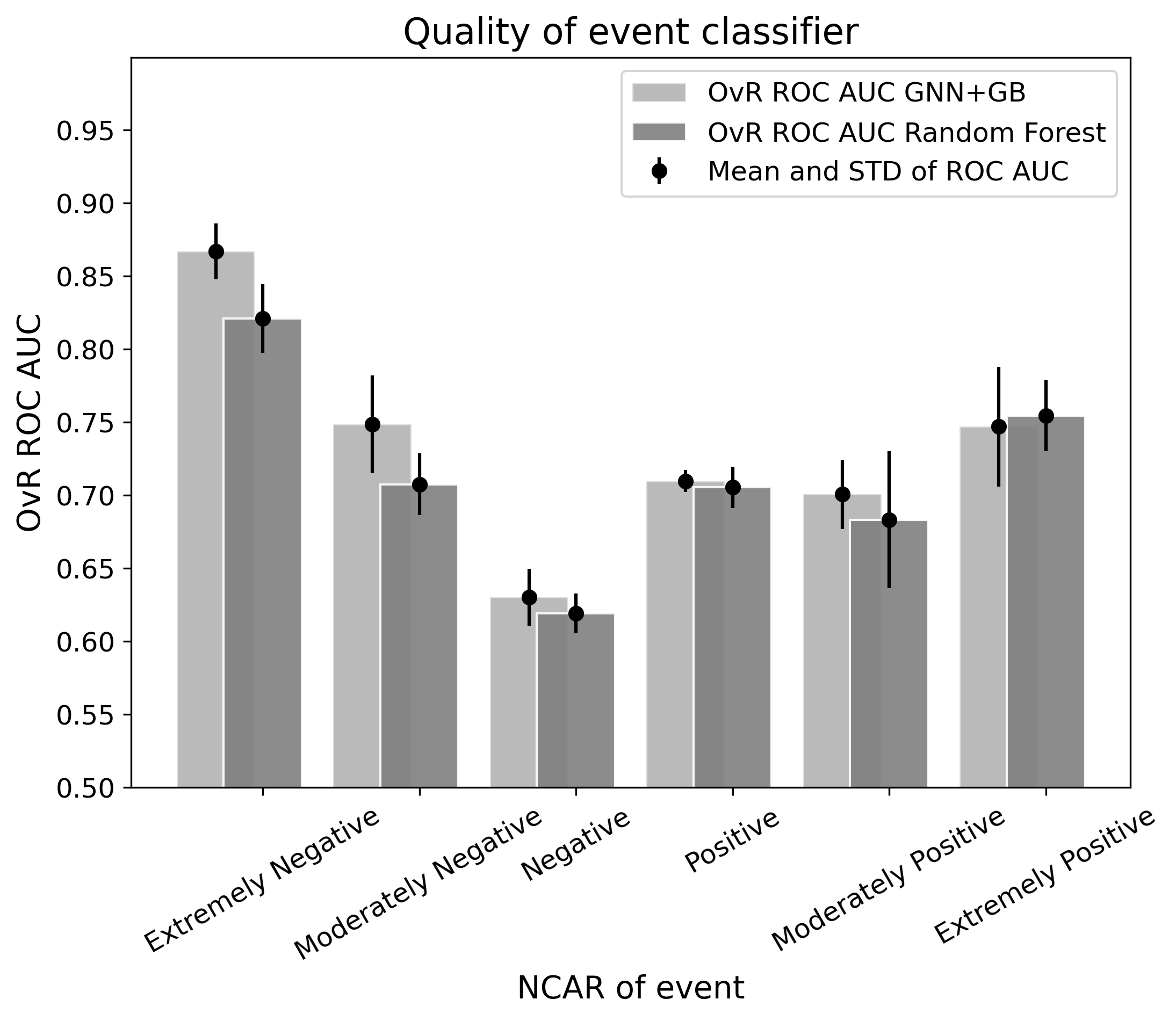}
        \caption{OvR ROC AUC metric for classification evaluation of GNN+GB and RF algorithms.} 
        \label{fig:Classifier results}
    \end{subfigure}
\label{fig:hist and AUC}
\caption{Samples distribution and classification metrics according to price change ranges.}
\end{figure}

Market conditions, company characteristics, announcement polarities and content, along with event relationships collected in one place make it possible to predict the price changes caused by the announcements. By setting our problem statement as a classification task, we categorize the stock price change values into six classes: Extremely Negative, Moderately Negative, Negative, Positive, Moderately Positive, and Extremely Positive. The number of classes is conditioned for three reasons. Firstly, it is important to know a sign of reaction. Secondly, we need to get a clear understanding of the impact amplitude. Thirdly, each class needs to be representative. Taking into consideration the indicated issues and the fact that our MAPE for the expected return evaluation is about 7\%, we define the range of price changes that belong to one class equal to 0.14. Figure \ref{fig:amp_distribution} shows that a small number of events cause price change greater than 0.28 in amplitude. Thus, due to the third reason, we combine announcement responses with price changes of more than 0.28 into one class. 

For the experiments, we split our dataset with the announcements in a stratified way by 10 times with the train and test subsets of 67\% and 33\%, respectively. As the classification model, we use the combination of GCN and GB. This allows us to improve the resulting classification quality by the combination of different types of data representation and the incorporation of the event interconnections. To figure out the advantage of adopting GCN, we calculate metrics for the case when prediction is made only with the classic machine learning model, namely, Random Forest (RF) classifier. RF shows the best result among all other classic machine learning models in the absence of GCN. One-vs-Rest (OvR) ROC AUC is taken as the main metric for the evaluation of model performance. The results are presented in Figure \ref{fig:Classifier results} and Table \ref{tab:Classifier results}.

\begin{table}[h!]
\begin{threeparttable}[b]
\centering
\caption{Events characteristics depending on the caused stock price change
and model performance metrics.}
\begin{tabular}[t]{lcccccccccc}
\toprule
\multirow{2}{*}{Class name\tnote{*}} & \textbf{Extremely} & \textbf{Moderately} & & & \textbf{Moderately} & \textbf{Extremely} \\
& \textbf{Negative} & \textbf{Negative} & \textbf{Negative} & \textbf{Positive} & \textbf{Positive} & \textbf{Positive} \\ 
Stock price change range & (-$\infty$, -0.28] & (-0.28, -0.14] & (-0.14, 0] & (0, 0.14] & (0.14, 0.28] & (0.28,+ $\infty$) \\
\midrule
% Class id & 0 & 1 & 2 & 3 & 4 & 5 \\
% Class description & Extremely Negative & Moderately Negative & Negative & Positive & Moderately Positive & Extremely Positive \\
Number of events & 211 & 189 & 599 & 478 & 110 & 67 \\
Positive events\tnote{**} & 72 & 106 & 421 & 366 & 83 & 57 \\
Negative events\tnote{**} & 139 & 83 & 178 & 112 & 27 & 10 \\
OvR ROC AUC for GCN+GB & $0.87 \pm 0.02$ & $0.77 \pm 0.03$ & $0.63 \pm 0.02$ & $0.71 \pm 0.01$ & $0.70 \pm 0.02$ & $0.75 \pm 0.04$ \\
OvR ROC AUC for RF & $0.82 \pm 0.02$ & $0.70 \pm 0.02$ & $0.62 \pm 0.01$ & $0.71 \pm 0.01$ & $0.68 \pm 0.05$ & $0.75 \pm 0.02$ \\
\bottomrule
\end{tabular}
\begin{tablenotes}
\item [*] According to the value of price change
\item [**] According to the sentiment polarity of the announcement
\end{tablenotes}
\label{tab:Classifier results} 
\end{threeparttable}
\end{table}

As it can be seen from Table \ref{tab:Classifier results}, the best model GCN+GB achieves OvR ROC AUC greater than 0.7 for all classes, excluding the Negative class. The model distinguishes the Extremely Negative class most accurately, demonstrating ROC AUC score of 0.87. The total weighted OvR ROC AUC for GCN+GB is equal to 0.71. RF shows the result of 0.69. Figure \ref{fig:Classifier results} proves the efficiency of using GCN, which integrates interaction between events due to construction specificity. There is an especially tangible impact on the results for the Extremely Negative, Moderately Negative, and Moderately Positive classes.     

\subsection{Analysis of feature importance}

\begin{figure}[h!]
        \centering
        \includegraphics[width=0.93\linewidth]{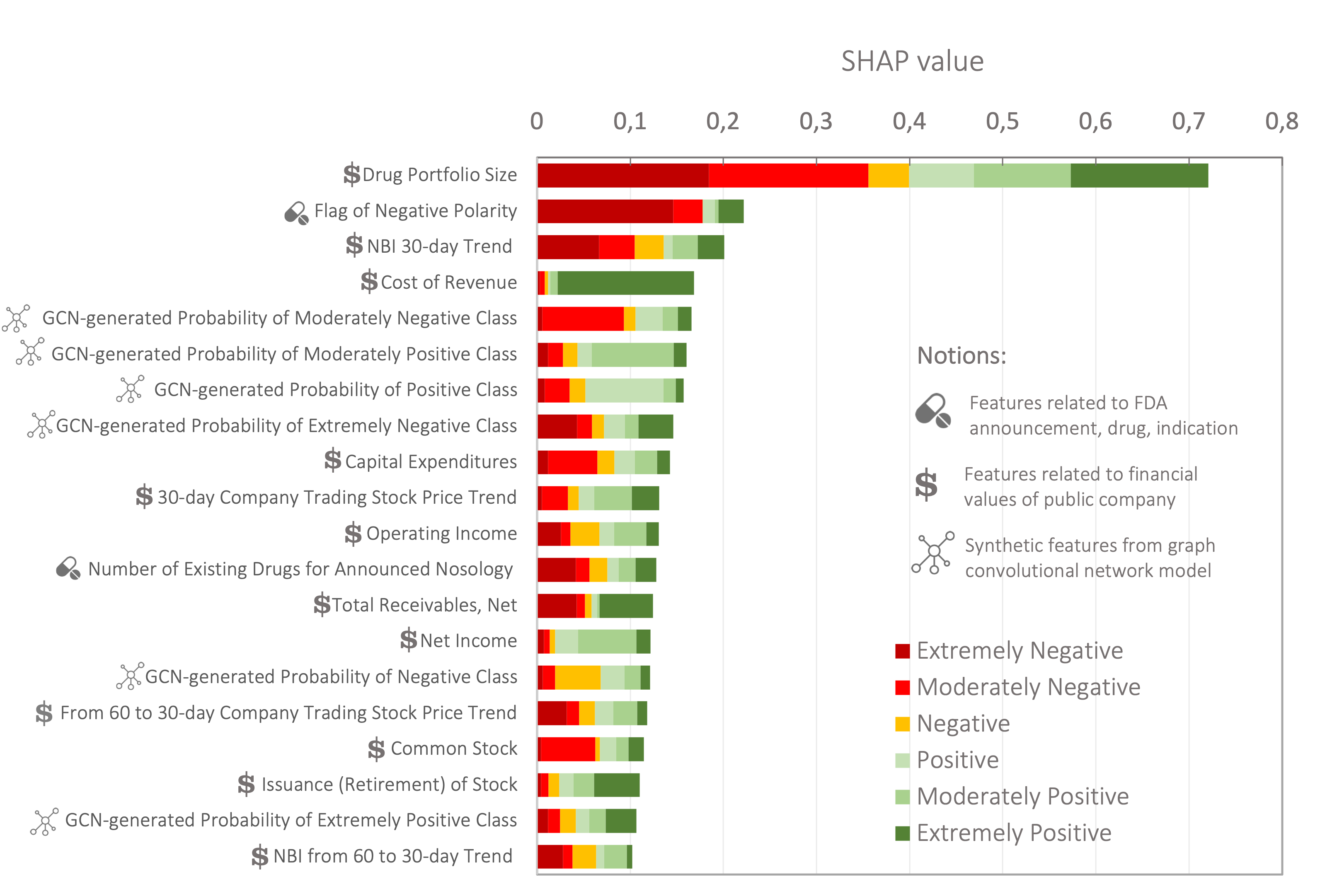}
        \caption{Top-20 of the most important features for Gradient Boosting Classifier according to SHAP.} 
        \label{fig:shap_fi}
\end{figure}

We manage to train the core classification model (GCN+GB) in a way that provides satisfactory performance metrics. For this reason, we can get a credible assessment of the importance of features utilized for prediction construction. We leverage the SHAP\cite{lundberg2017SHAP} method for this purpose. Feature importance is estimated when using GB, which takes constructed feature space and the class probabilities from GCN as input. The resulting distribution is provided in Figure \ref{fig:shap_fi}.

The "Drug Portfolio Size", "Flag of Negative Polarity", and "NBI 30-day Trend" features occupy the first three positions in terms of importance. Notably, all the features generated by GCN are located in Top-20. Many top features are associated with the indicators from reports or statistics on the trend data. "Number of Existing Drugs for Announced Nosology" stands out over other features, as it is designed using hybrid information from medicine and the announcement. Interestingly, there are no features in the Top-20 that indicate 
the clinical trial phase when the announcement is released.

\section{Discussion} \label{sec:discussion}

In this work, we provide a statistically justified and trustworthy framework for operation with the pharma stock market and its reaction to the trial result announcements, handling one of the biggest datasets of \numnews releases by \numcomp pharmaceutical companies. Examining available data in detail, we unveil several noteworthy behavioral patterns in the stock market. In particular, from Figure \ref{fig:distributions_3sub}, we can see that the negative events lead to a greater and more definitive impact on the share prices than the positive ones. Specifically, the negative events cause more significant asymmetry in the price responses, expressed in sharp financial losses in some cases. This kind of effect is reported in many studies and is usually explained by overconfidence in future success\cite{kim2016ceo}. Such overconfidence implies the presence of positive expectations in the recent share prices. Thereby, the prices should not change much if the results are indeed positive. Also, a stronger reaction to negative announcements can be substantiated by the ambiguity of future state of affairs after positive events in comparison with the certainty after negative ones\cite{sharma2004linking}. Negative trial results often signify that the treatment under development will not be forthcoming, consequently inflicting a financial hit. Furthermore, we find out that the drug portfolio size of the company is a key characteristic governing the price reaction. In short, the more diversified the company is, the less harmful effect from the negative trial test announcements may be expected. Importantly, we manage to identify the precise risk boundaries. As it can be seen from Figure \ref{fig:portfolio_size}, the companies with a portfolio size varying from 1 to about 5 products experience a strong sensitivity to the announcements. In reviewing the resulting values of the core classifier metrics, we point out that the model shows bad quality for the Negative class ($(-0.14,0]$ interval of $NCAR_{\postnewsperiod}$) in comparison with other classes, although it contains the highest number of samples. We suppose that firstly it can be connected with the correction of stock price because of profit fixation. Secondly, the error of expected return prediction may also lead to prediction inaccuracies. Applying the model interpretation technique, we extract feature importances. Their distribution confirms the significance of attributes responsible for the company drug portfolio size, the announcement sentiment polarity, and the GCN-generated probabilities, disclosing at the same time the importance of the trend information or such specific feature as the number of drugs in nosology announcements.

Nevertheless, the developed framework for predicting the stock price response implies several assumptions and limitations. The basic assumption is that the stock prices incorporate all relevant information available to market traders. In this case, the stock price is instantaneously affected by newly revealed information related to the examined company\cite{mcwilliams1997event}. Next, we assume there is no information leakage in a market before the release takes place. It means we set an official announcement as a start point for the market reaction. It is worth noting that the official FDA announcements could be published at a time when the preliminary results are publicly available. In fact, the stock price of publicly traded pharmaceutical companies tends to react to the clinical trial results before official releases\cite{rothenstein2011company}. Finally, we isolate the effect of clinical result announcements from the other events. This assumption is the most critical because there are definitely other factors impacting the stock prices (e.g., declaration of dividends, financial report, new product announcement, merging with another company, etc.). Nevertheless, we believe that clinical trials have one of the most pronounced impacts on the producer's financial state. At the same time, we attempt to account for other non-visible events by tracking the trading volume. Usually, the most active moments of trading correspond to the major events that are not necessarily associated with the company. 

\section{Conclusions} \label{sec:conclusions}

In this study, we quantitatively relate clinical trial announcements with the market value change of pharmaceutical companies. In particular, we construct a statistically justified predictive framework for price change evaluation and prove a principal presence of forecast potential for future FDA announcement impacts by achieving the total weighted ROC AUC score greater than 0.7 for historical data. Notably, we are able to predict Extremely Negative, Moderately Negative, and Extremely Positive classes most precisely with 0.87, 0.77, and 0.75 ROC AUC scores, respectively. Since we analyze one of the biggest announcement datasets, we extract reliable relationships between the company background and the price change peculiarities. Our research offers insights into the behavior of the stock prices and can constitute a solid basis for the further more sophisticated analysis of the pharma market.

Summing up, our findings can be helpful for people who deal with the pharma market and want to make more rational strategic investment decisions, hedging their financial risks. In addition, this work contributes to the event studies and demonstrates the direct application of machine learning tools in the industry.  

\bibliography{sample}

\begin{thebibliography}{10}
\bibitem{Aravive2017}
\bibinfo{title}{Versartis announces phase 3 velocity trial of somavaratan in
  pediatric growth hormone deficiency did not meet primary endpoint}.
\end{thebibliography}

\section*{Competing interests}
The authors declare no competing interests.

\section*{Author contributions statement}

Drafting of the manuscript: S.B., A.K., E.K., L.Z. Design and modeling: S.B., A.K. Data collection and analysis: S.B., A.K.

\section*{Additional information}

\textbf{Correspondence} and requests for materials should be addressed to S.B.

\section*{Supplementary materials}

\subsection*{Post-announcement period calculation}

\begin{figure}[h!]
\centering
    \includegraphics[width = 0.65\linewidth]{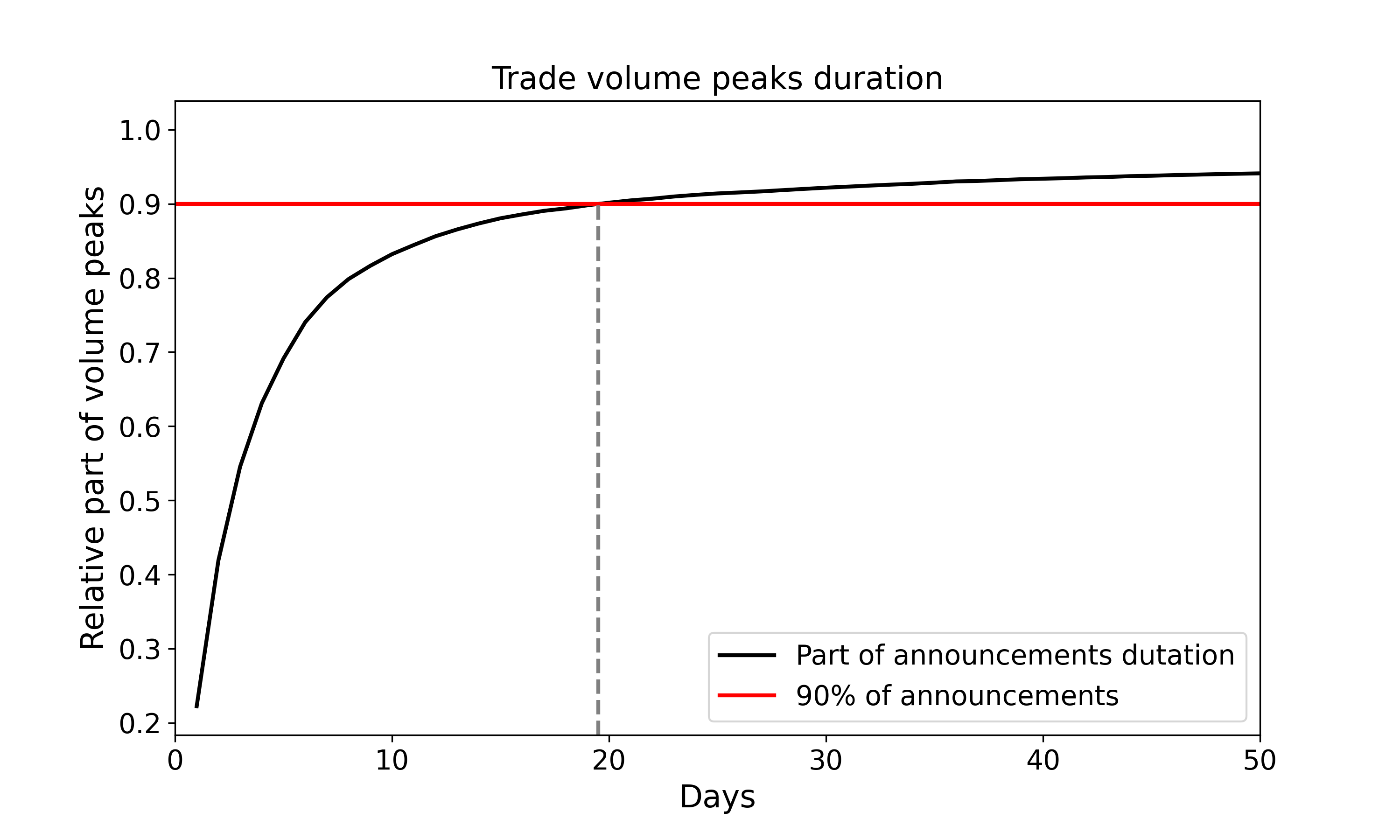}
    \caption{The correspondence between the share of considered trading volume peaks with the peak duration. Redline is the 90\% of the total number of peaks. }
    \label{fig:peaks_duration}
\end{figure}

To estimate the duration of the announcement effect, we analyze trading volume peaks in the companies' history. These peaks do not exclusively mean the reaction to the trial results but are rather associated with all events related to the company. We calculate the duration in days of all available trading volume peaks. The results are presented in Figure ~\ref{fig:peaks_duration}. As we can see, the duration of 90\% peaks constitutes 20 trade days. Thus, we take 20 days as the post-announcement time period we are interested in.

% To find out post-announcement time period, we calculated the  post-announcement time for trading volume rate to get lower 2 median trading volumes and cut off the 90\% percentile. 

\subsection*{Other relationships between announcement impact and company characteristics}

\begin{figure}[h!]
    \centering
    \includegraphics[width=0.45\linewidth]{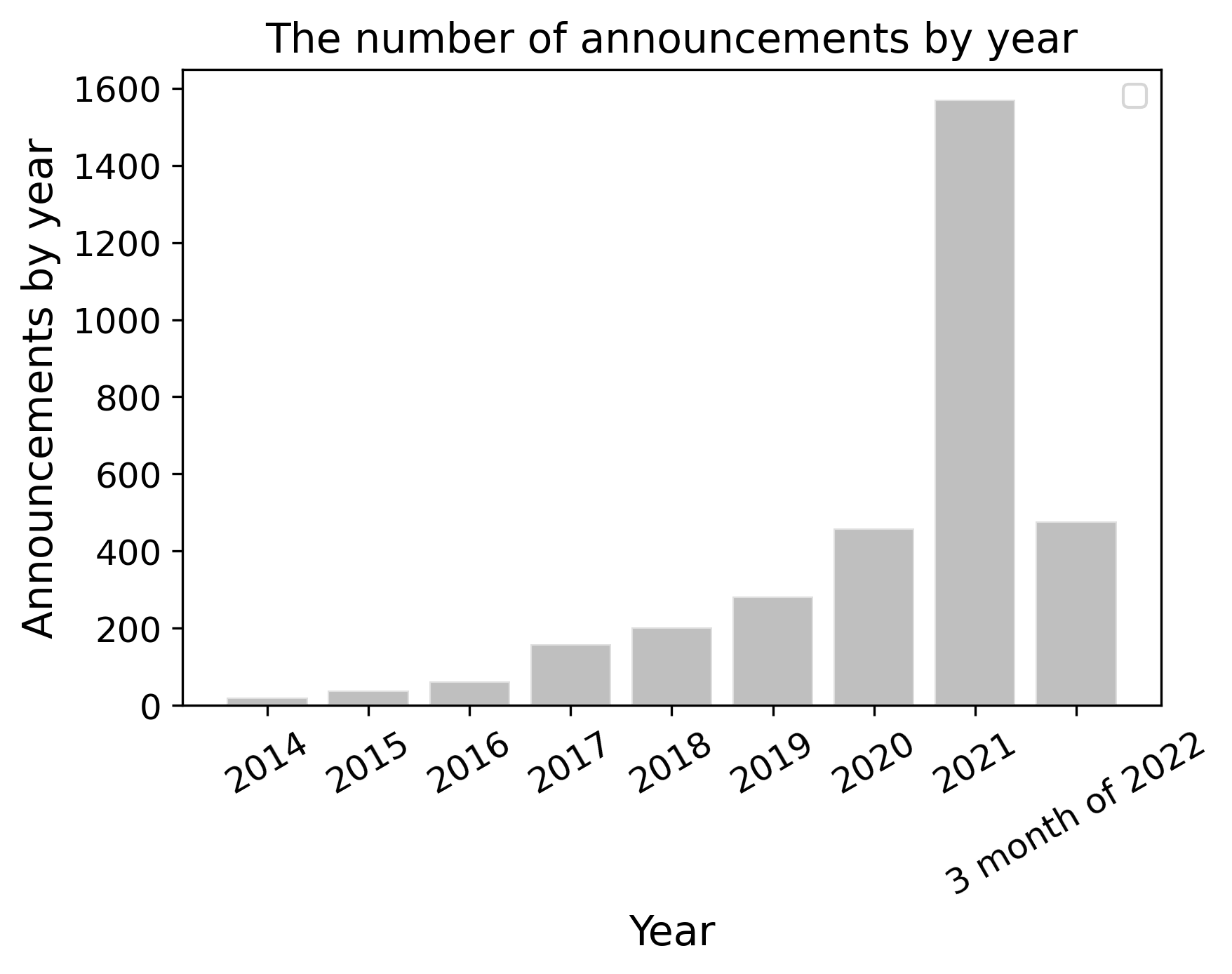}
    \caption{The number of FDA announcements of public companies in different years.}
    \label{fig:announcements_by_year}
\end{figure}

The dependence of the number of FDA news announcements on the year for the public companies is given in Figure \ref{fig:announcements_by_year}. It demonstrates the increase in the number of FDA announcements in open sources by approximately ten times from 2017 to 2021.  

\begin{figure}[h!]
    \centering
    \includegraphics[width = \linewidth]{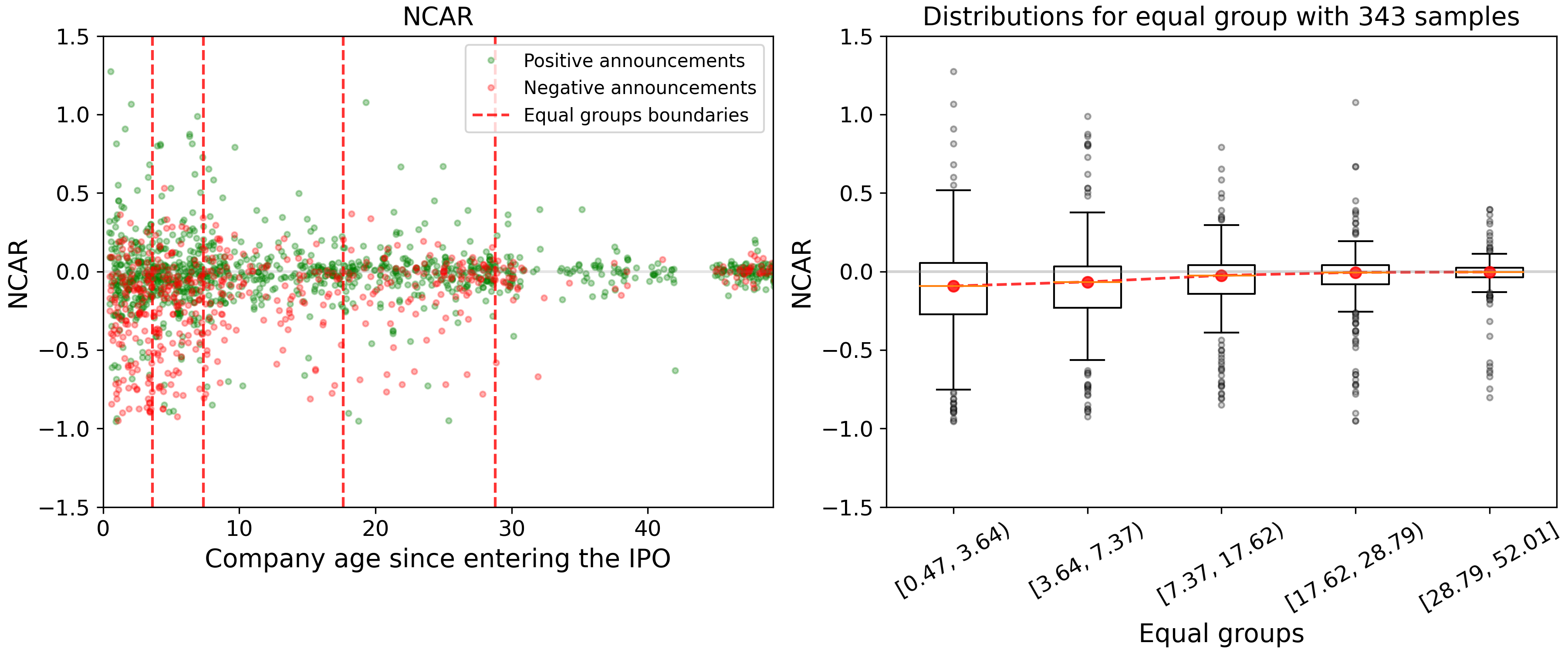}
    \caption{Dependence of the stock price changes on the company age. In the left part, the red dashed lines divide announcements into equal groups (with the same number of announcements inside). In the right part, the red dashed line goes through the median values of each group's stock price changes.}
    \label{fig:123}
\end{figure}

\begin{figure}[h!]
    \centering
    \includegraphics[width = \linewidth]{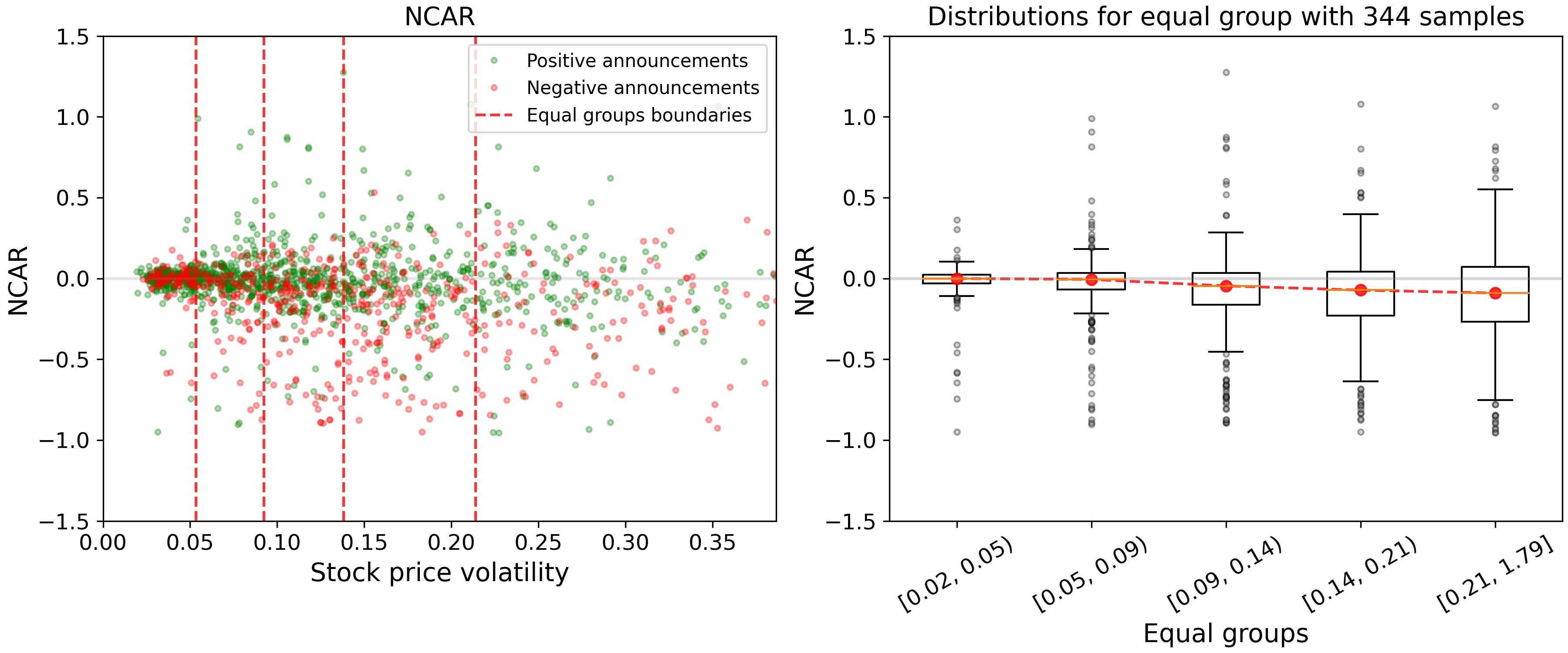}
    \caption{Dependence of the stock price changes on the company volatility. In the left part, the red dashed lines divide announcements into equal groups (with the same number of announcements inside). In the right part, the red dashed line goes through the median values of each group's stock price changes.}
    \label{fig:124}
\end{figure}

\begin{figure}[h!]
    \centering
    \includegraphics[width = \linewidth]{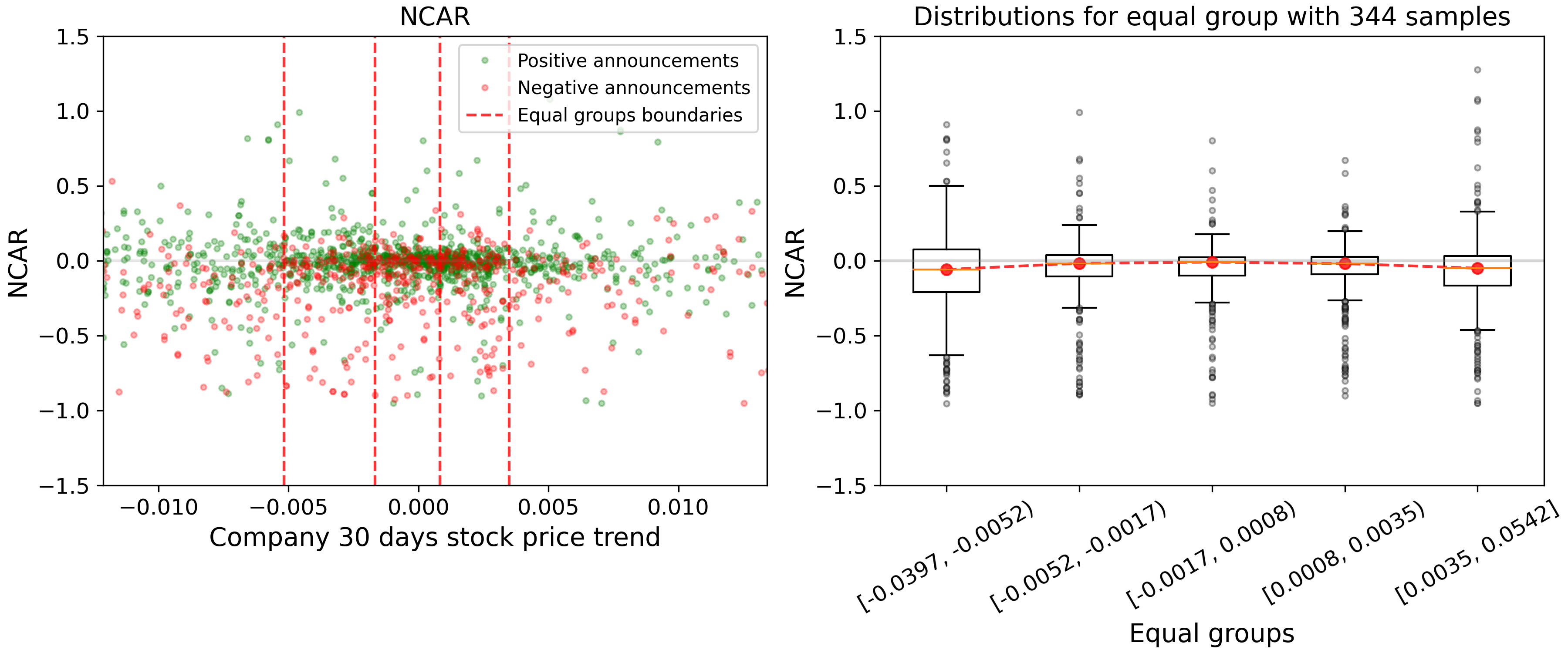}
    \caption{Dependence of the stock price changes on the 30-day trend before the announcement. In the left part, the red dashed lines divide announcements into equal groups (with the same number of announcements inside). In the right part, the red dashed line goes through the median values of each group's stock price changes.}
    \label{fig:125}
\end{figure}

Figures \ref{fig:123}, \ref{fig:124} and \ref{fig:125} show the dependence of stock price changes caused by announcements on the company age since entering the IPO at the announcement moment, historical volatility, and the company's 30-day stock price trend. Volatility is defined as the relation of the standard deviation of stock price to the median stock price in a 200-day period before the announcement. We can conclude that the younger company is, the more sensitive it is to the announcements. Greater intrinsic company volatility results in more extreme price changes. In addition, if the company trend of a stock price is about zero in the pre-event period, then the induced price change is not significant.

\subsection*{Mismatch analysis between announcement polarities and price change}  

\begin{figure}[h!]
    \centering
    \includegraphics[width=0.95\linewidth]{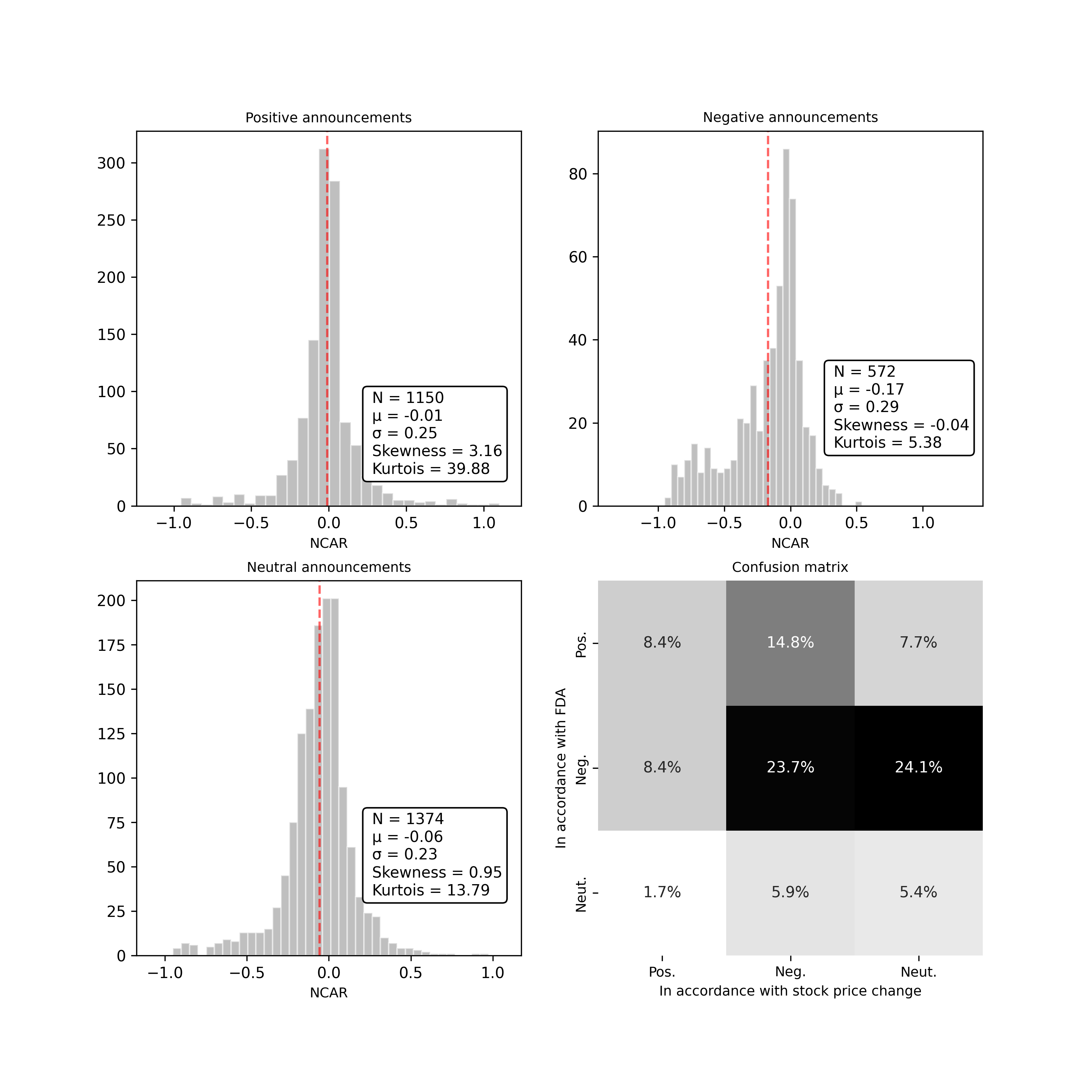}
    \caption{The price change distribution ($NCAR_{\postnewsperiod}$) for positive (upper left), negative (upper right), neutral (lower left) FDA news with statistical parameters provided in the boxes (number of events $N$, mean value $\mu$, standard deviation $\sigma$, skewness and kurtois coefficients). The mean value is depicted with the red vertical dashed line. The confusion matrix (lower right) of mismatch between sentiments in accordance with FDA announcements and polarities in accordance with the stock price change ($NCAR_{\postnewsperiod}$).} 
    \label{fig:distributions}
\end{figure}

To realize the mismatch between events sentiment in accordance with FDA announcements and events sentiment in accordance with the actual price change, we build a confusion matrix shown in Figure \ref{fig:distributions}. The sentiment polarity derived from stock price change, namely $NCAR_{\postnewsperiod}$, is defined as follows: the event is set to be positive or negative one in accordance with $NCAR_{\postnewsperiod}$ if its price change is more than $\sigma_{neut}/2 = 11.5\%$ ($\sigma_{neut}$ - standard deviation of $NCAR_{\postnewsperiod}$ for neutral announcements) or less than $-\sigma_{neut}/2$, respectively. If the price change lies within $[-\sigma_{neut}/2; +\sigma_{neut}/2]$, then we define such event as a neutral one in accordance with $NCAR_{\postnewsperiod}$. As a result, the highest rates in the confusion matrix constitute 23.7\% and 24.1\% and relate to the cases in which negative announcements have negative or neutral price changes.

\subsection*{Statement on computational resources and environmental impact}  

We used a NVIDIA GeForce RTX 2070 SUPER GPU and NVIDIA A100 80GB PCIe GPU to train the classifiers and the BERT model, correspondingly.  

This work contributed 1.34 kg and 6 g of equivalent $CO_{2}$ emissions during the classifiers and BERT training, respectively. The carbon emissions information was generated using the open-source library eco2AI\cite{eco2ai}.

\end{document}